\documentclass[journal=jacsat,manuscript=article]{achemso}

\usepackage[version=3]{mhchem} % Formula subscripts using \ce{}

\author{Johan Klarbring}
\email{johan.klarbring@liu.se}
\affiliation{% 
Department of Materials, Imperial College London, Exhibition Road, London SW7 2AZ, United Kingdom
}%
\alsoaffiliation{% 
Department of Physics, Chemistry and Biology (IFM),
Link\"{o}ping University, SE-581 83, Link\"{o}ping, Sweden
}%
\author{Aron Walsh}
\affiliation{% 
Department of Materials, Imperial College London, Exhibition Road, London SW7 2AZ, United Kingdom
}%

\title{Na Vacancy Driven Phase Transformation and Fast Ion Conduction in W-doped Na$_3$SbS$_4$ from Machine Learning Force Fields \\
       }
\begin{document}

\begin{tocentry}

\includegraphics{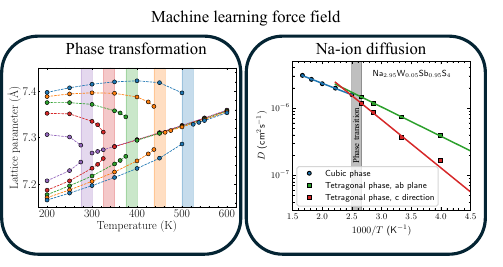}

\end{tocentry}

\begin{abstract}
Solid-state sodium batteries require effective electrolytes that conduct at room temperature. The \ce{Na3PnCh4} (Pn = P, Sb; Ch = S, Se) family have been studied for their high Na ion conductivity. The population of Na vacancies, which mediate ion diffusion in these materials, can be enhanced through aliovalent doping on the pnictogen site. To probe the microscopic role of extrinsic doping, and its impact on diffusion and phase stability, we trained a machine learning force field for Na$_{3-x}$W$_{x}$Sb$_{1-x}$S$_4$ based on an equivariant graph neural network. Analysis of large-scale molecular dynamics trajectories shows that an increased Na vacancy population stabilises the global cubic phase at lower temperatures with enhanced Na ion diffusion, and that the explicit role of the substitutional W dopants is limited. In the global cubic phase we observe large and long-lived  deviations of atoms from the averaged symmetry, echoing recent experimental suggestions.  Evidence of correlated Na ion diffusion is also presented that underpins the suggested superionic nature of these materials. 
\end{abstract}

\section{\label{sec:Intro}Introduction}

As the development of solid-state lithium-ion batteries is maturing,\cite{Kim2015,Janek2016,Zhao2020} an increased emphasis is being placed on alternative battery chemistries to reduce future lithium demand for energy storage technologies. Beyond Li, Na-ion batteries have distinct advantages in that Na is cheaper and more abundant\cite{Zhao2018}. 
Among promising candidate materials, the class Na$_3$PnCh$_4$, where Pn = P or Sb and Ch = S or Se, stands out \cite{Jia2021}. Na$_3$PS$_4$ was demonstrated as a solid-state electrolyte in 2012 by Hayashi \textit{et al.} \cite{Hayashi2012}, and a host of studies have followed, attempting to understand and optimize materials in this class. 

It has since been realised that Na$^{+}$ vacancies are key to the high Na$^{+}$-diffusivity. Indeed, samples prepared with low intrinsic Na$^{+}$-vacancy concentrations are poor ionic conductors, while samples with higher defect concentrations show much higher conductivities \cite{Zhang_2016,Krauskopf_2018,Maus_2023}. Along these lines, several studies have identified aliovalent substitutional doping as an effective way to introduce charge-compensating Na$^{+}$-vacancies, and thus boost the ionic conductivity in these materials. In particular, doping W$^{+6}$ on the Pn$^{+5}$-site has been demonstrated to effectively increase the ionic conductivity and Na$_{3-x}$W$_{x}$Sb$_{1-x}$S$_4$, for x $\sim$ 10-12 \% shows among the highest room temperature (RT) Na-ion conductivity of any solid-state material to date\cite{Hayashi_2019,Fuchs_2019}.

Materials in the Na$_3$PnCh$_4$ class crystallize in a tetragonal structure (space group $P\bar{4}2_{1}c$) at low temperatures and transform to a cubic phase (space group $I\bar{4}3m$) on heating. In addition to boosting the ionic conductivity, substitutional doping, has been shown to have a large influence on the tetragonal to cubic phase transformation temperature, $T_C$. Indeed, while pristine Na$_3$SbS$_4$ stays tetragonal up to at least $\sim$ 440 K \cite{Zhang2018}, Na$_{2.9}$W$_{0.1}$Sb$_{0.9}$S$_4$ transforms below RT \cite{Maus_2023}.

An interesting feature of these materials is that different samples show either a tetragonal or cubic average structure depending on synthesis route \cite{Krauskopf_2018}, while local structural probes, such as eg. pair distribution function (PDF) measurements, show that the local structure is better described by the tetragonal phase, even in samples where the long-range averaged structure is cubic \cite{Krauskopf_2018}.  In particular, recent work on W-doped Na$_{3}$SbS$_4$ \cite{Maus_2023} showed that 10\% W-substitution yields an averaged cubic phase at RT, while PDF, Raman and nuclear magnetic resonance (NMR) measurements hint at a local structure of lower symmetry. This type of discrepancy between long-range global- and local symmetry appears to be a feature of many modern energy materials, eg.\ the halide perovskites\cite{weadock2023nature,baldwin2024dynamic}.

The functionality of these materials is the result of an intricate interplay between doping, local- and average structure, phase transformations and ionic diffusion, and the separate effects of each of these phenomena are not fully understood. Atomistic modeling offers an attractive route to address these questions, and several density functional theory (DFT) based studies have been performed in recent years. Nevertheless, the inherently high computational cost of DFT prohibits simulations from covering the range and time-scales required to properly investigate the phenomena mentioned above. Towards this end, classical molecular dynamics (MD) simulations can be leveraged, as for instance done by Sau $\textit{et al}$ to study the ion conduction in Na$_3$PS$_4$\cite{Sau2020}. The force-fields used in classical MD studies, however, are often not accurate enough and, in particular, often suffer from transferability issues between different systems and across dopant ranges.

Machine-learned force-fields (MLFFs) offer a potential route to overcome these issues\cite{Ko2023}. Indeed, rapid progress has been made in recent years in the design of accurate and computationally efficient machine-learning architectures. These modern MLFFs are being applied to study progressively more complex problems, including ionic diffusion in solid-state electrolytes, and phase transformations in complex energy materials.  Nevertheless, their accuracy in describing phase-transformations and diffusion, and their interplay, in substitutionally doped solid-state electrolyte materials, remains an open question.

In this work, we construct an MLFF based on the Allegro \cite{Musaelian2023,Kozinsky2023} architecture, an equivariant graph neural network, capable of running large-scale MD simulations and accurately describing and providing physical insight into the intricate interplay between substitutional W-doping, structural phase transformations and diffusion in Na$_{3-x}$W$_x$Sb$_{1-x}$S$_4$. Using our trained MLFF, we show that W-doping, provided that it is accompanied by charge-compensating Na-vacancies, results in a monotonous decrease in the cubic-to-tetragonal phase transformation temperature, in full agreement with available experimental data. We then show that this reduction in $T_C$ is an effect of the Na-vacancies, rather than the W-dopants.  We further show that our model can reproduce the Na-ion diffusion in fair agreement with experimental data, and that, again, W-dopants have little effect on the diffusion other than introducing Na-vacanies. We also explicitly show, through calculation of the Haven ratio, that the Na-ion diffusion in these systems is correlated. In summary, we demonstrate that a carefully constructed MLFF can describe diffusion and phase transformations in prospective doped Na-ion electrolyte materials. 
 
\section{\label{sec:Methods}Methodology}

\subsection{Allegro Machine Learning Force Field construction}
Allegro \cite{Musaelian2023} is a recently developed equivariant graph neural network (GNN) potential. While many MLFF architectures ensure translational and rotational invariance of the predictions by using invariant scalar descriptors based on eg. distances, angles or dihedrals, the equivariant GNNs act directly on displacement vectors in a symmetry-respecting way\cite{Batzner2022}. This can result in more accurate, stable and data efficient models. Different to several equivariant GNNs \cite{Batzner2022,Batatia2022mace}, which rely on message passing, Allegro is strictly local, which allows for efficient parallelization and makes simulations of very large system sizes possible \cite{Musaelian2023,Kozinsky2023}.

To generate a robust training set, we utilize a two step procedure. First, we generate a set of configurations using an on-the-fly learning procedure implemented in VASP \cite{Jinnouchi2019,Jinnouchi2019_2} where a Gaussian approximation potential (GAP)-style \cite{Klawohn2023} potential is re-fit to data picked out during and MD run based on a Bayesian error prediction. These runs are performed on 128 atom supercells corresponding to $2\times2\times2$ expansions of the conventional cubic unit cell. We do separate runs for a few different Sb-W configurations ranging from 0 to 4 W-dopants (up to 25 atomic \%) with 1 charge-compensating Na-vacancy introduced per W dopant. Using this training set we fit a preliminary Allegro MLFF. The purpose of this initial model is purely to allow us to run cheap and stable MD simulations for a range of W-dopant concentrations. Second, using this preliminary force-field, we perform MD simulations using the atomic simulation environment (ASE) \cite{HjorthLarsen2017} in $2\times2\times4$ supercells for a total of 11 different W/Sb configurations, with W-concentrations ranging from 0 to 25 \%. These configurations cover both highly clustered and highly separated distributions of W dopants. Each W/Sb configuration is ran for 200 ps at 700 K and 200 ps at 200 K. We then pick out a total of 2861 configurations from these MD runs and run single-point DFT calculations on them to make up the final training and validation sets.

Our Allegro model used a 6.5 Å radial cutoff and 2 layers. We used 32 tensor features with $l_{max}=2$ and full O(3) symmetry. The 2-body latent multi-layer perceptron (MLP) and later latent MLP had dimensions [64, 128, 256, 512] and [512], respectively and SiLU non-linearities. The final edge-energy MLP had dimensions [128] without non-linearity.  Interatomic distances were embedded using trainable Bessel functions. 
The training and validation set consisted of 2289 and 572 structures, respectively, and were reshuffled after each epoch during training. The loss function was equally weighted between Allegros per atom energy, force and stress terms. The training used the pytorch \cite{pytorch_2019} Adam optimizer and ran for 1121 epochs using a batch size of 5 and a learning rate of 0.001.

A test set was generated using the final Allegro model by running MD using $2\times2\times4$ supercells at 600, 400 and 200 K for 10 ps each using 6 different W/Sb configurations ranging from 0 to 18.75 \%. These W/Sb configurations were generated independently from the training set. The test set contained 180 structures. The final model achieves root mean squared errors (RMSE) on this test set of 0.38 meV/atom, 28 meV/Å and 0.18 kbar, for energies, force components and stress components, respectively. See SI Fig.\ 1 for parity and error distribution plots. The model is further validated versus DFT by comparing nudged elastic band (NEB) diffusion barriers, phonon dispersions and soft-mode potential energy surfaces (PES), achieving satisfactory accuracy in all cases, see SI.

The final training and validation and test set as well as the final trained model is available at \texttt{10.5281/zenodo.10891472}.

Simulations using the final MLFF were run on stoichiometric \ce{Na3SbS4} and the W-doped system with charge compensating Na-vacancies, Na$_{3-x}$W$_x$Sb$_{1-x}$S$_4$. In addition, we also performed simulations on Na-deficient systems, \ce{Na_{3-x}SbS4}, and on W-doped systems with no compensating Na-vacancies, \ce{Na_{3}Sb_{1-x}W_xS4}, where the charge compensation is implicit. Here the potential energy surface described charged sodium vacancies/W-dopants, but without the explicit presence of compensating W dopants/Na-vacancies. This is feasible since the potential is short-ranged and local, and the training set contains large-enough regions with and without dopants in proximity of the Na vacancies.

\subsection{Density Functional Theory Calculations}
All DFT calculations used for the training, validation and tests sets were performed using VASP \cite{kresse1996,kresse1996_2,kresse1999}, within the Projector Augmented Wave formalism \cite{blochl1994} and the r$^2$SCAN \cite{Furness2020} exchange-correlation functional. For the final training set we used a cutoff energy of 520 eV and a $2\times2\times1$ Monkhorst-Pack \textit{k}-point grid for the $2\times2\times4$ (256 atom) supercells. The threshold for the electronic self-consistent field iterations was set to 10$^{-6}$ eV and a 50 meV Gaussian smearing was applied to the electronic occupancies. We used the default recommended VASP PBE-PAW potentials, labeled 'W\_sv', 'Sb', 'Na\_pv' and 'S' for the corresponding elements.

\subsection{Molecular Dynamics Simulations}
All ML-MD simulations were performed using LAMMPS \cite{Thompson_2022} with the pair\_allegro \cite{pair_allegro} patch. We used a 2 fs timestep, and a Nose-Hoover thermostat and barostat using time constants of 0.2 ps and 2 ps, respectively. W-dopants were distributed in the supercells on Sb sites using the special quasi random structure (SQS) methodology\cite{Zunger1990}, as implemented in icet \cite{angqvist2019}, and the Na-vacancies were introduced by randomly removing Na atoms. 

To determine phase transformation temperatures we use the following procedure. First, starting from a 12$\times$12$\times$12 conventional-cubic supercells, NPT MD simulations with a fully flexible cell were performed on a coarse temperature grid (50 K spacing) between 200 and 600 K for 200 ps. Based on the behaviour of the lattice parameters after an initial equilibration period, each of these temperature points were then assigned either to the tetragonal phase, cubic phase or a mixture. In the temperature region between the highest temperature that can be identified to be in the tetragonal phase, and the lowest temperature which can be clearly identified to be in the cubic phase, we ran two sets of new simulations on a 12.5 K spaced grid. One of these sets used the positions and velocities from the tetragonal endpoint and the other those from the cubic endpoint. These simulations were ran for at least 400 ps, as needed. The temperature points on this more tightly spaced grid where then assigned to one of the phases when the simulations from both staring points could be identified (after an equilibration period) to be in the same phase. Typically, at one temperature, we saw long timescale shifts between the tetragonal and cubic phases (see SI Fig.\ 5 for an example). These shifts between cubic and tetragonal lattice parameters is an indication that the system is in the vicinity of the phase transformation and that there is a (weak) first-order nature associated with the transition, at least for the 12$\times$12$\times$12 supercell sizes used in this procedure. In the end, this procedure yields phase transformation temperatures with a $\pm$ 12.5 K associated uncertainty, which is accurate enough for our purposes. 

The self diffusion coefficients, $D$ and the charge diffusion coefficient, $D_\sigma$, were obtained using 'kinisi' \cite{McCluskey2023}. For the self diffusion coefficient, we used the averaged lattice parameters extracted from the NPT runs, and  performed at least 0.5 ns of equilibration in the canonical (NVT) ensemble  using 8$\times$8$\times$8 supercells, followed by at least 0.5 ns in the microcanonical (NVE) ensemble using from which $D$ was extracted.

The Haven ratio $H_R$, was obtained as $H_R = D/D_\sigma$. As $D_\sigma$ requires significantly longer trajectories to converge, compared to $D$, we used 4$\times$4$\times$4 supercells to allow for longer simulation time. At each temperature we performed 4 separate runs, these runs were equilibrated for 0.5 ns in the NVT ensemble, followed by long runs in the NVE ensemble. In total, 48 ns of NVE dynamics was used to extract $D_\sigma$ at each temperature. 

The self diffusion coefficients agree closely between 4$\times$4$\times$4 and 8$\times$8$\times$8 supercells, as shown in SI Fig.\ 8

\subsection{Phonon Dispersion Relations}
We calculated harmonic phonon dispersion relations of cubic \ce{Na3SbS4} with both r$^2$SCAN and Allegro using 2$\times$2$\times$2 supercells. We generate small displacements with amplitudes 0.02 Å using phonopy\cite{phonopy1,phonopy2} and then extracted the interatomic force constants, phonon dispersion relations and density of states (DOS) using the temperature dependent effective potential (TDEP) package \cite{Hellman2013,Knoop2024}.

\section{\label{sec:Results}Results}
\begin{figure*} 
\centering
\includegraphics[width=\linewidth, keepaspectratio=true]{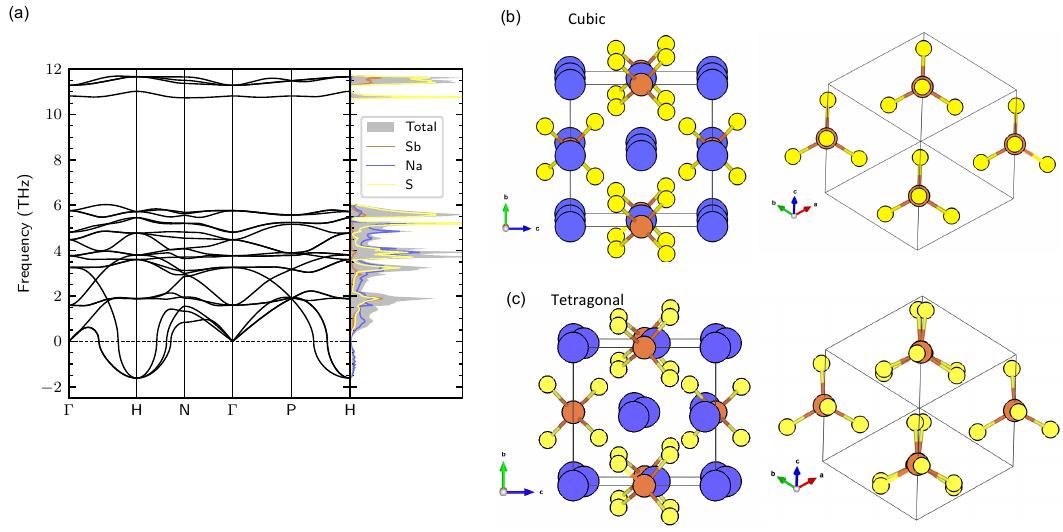}
\caption{ (a) Phonon dispersion relations in cubic Na$_3$SbS$_4$. (b) Cubic  and (c) Tetragonal polymorphs of Na$_3$SbS$_4$.  The right panels of (b) and (c) show a view along the [111] direction (Na-ions are removed for clarity), highlighting the relative tilt of sequential SbS$_4$ tetrahedra away from perfect alignment in the tetragonal phase.}
\label{fig:fig1}
\end{figure*}

\subsection{\label{sec:phase_trans}Structure and Phase Transformation}

\begin{figure}
\centering
\includegraphics[width=0.5\linewidth, keepaspectratio=true]{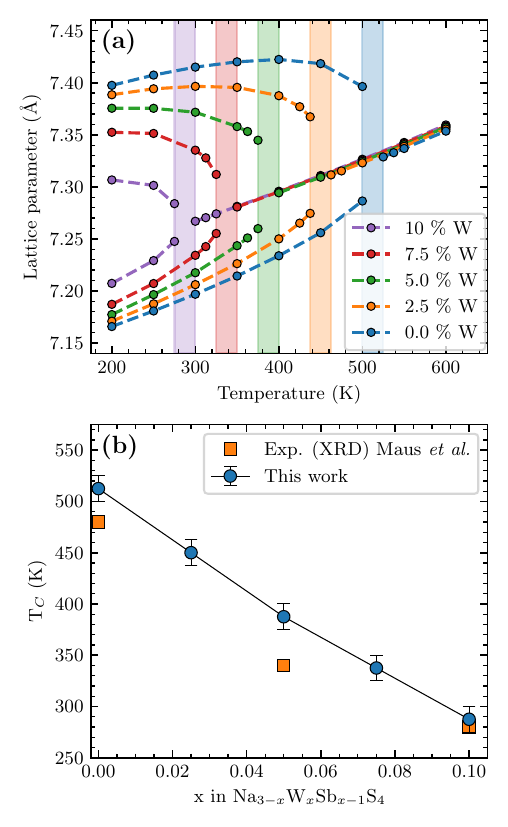}
\caption{(a) Lattice parameters vs temperature for Na$_{3-x}$W$_x$Sb$_{1-x}$S$_4$ for $x$ in 0-0.1. The shaded regions indicate the phase transformation temperatures. (b) Cubic-tetragonal phase transformation temperature, $T_C$, as a function of $x$ in Na$_{3-x}$W$_x$Sb$_{1-x}$S$_4$, compared to experiment (XRD) from Ref.\ \cite{Maus_2023}.}
\label{fig:phase_trans}
\end{figure}

Fig. \ref{fig:fig1} (a) shows the calculated phonon dispersion of cubic Na$_3$SbS$_4$ with its characteristic imaginary phonon mode at the H-point. This mode has been connected to the cubic-to-tetragonal phase transformation, as is the case for several compounds from this class of materials\cite{Brenner2022}. Indeed, the eigendisplacements of this mode largely overlaps with those that transforms the system from cubic to the tetragonal phase\cite{Gupta2021}. The tetragonal phase is obtained from the cubic phase by (1) a shuffling of Na-ions along the [001] directions, and (2) a rotation of the SbS$_4$ tetrahedra around [111], as indicated in Fig.\ \ref{fig:fig1} (c). There is also an accompanying elongation of the structure along the $c$-axis, yielding $ c/a > 1$ in the tetragonal phase.

To probe the tetragonal to cubic transformation we perform large-scale ($\sim$ 27.6k atoms) NpT MD simulations for a range of  different temperatures. The blue markers in Fig. \ref{fig:phase_trans} show the resulting averaged lattice parameters as a function of temperature. Below 500 K Na$_3$SbS$_4$ is in the tetragonal phase with the $a=b$ different from the $c$ lattice parameter, while above 525 K, $a=b=c$, and the system is in the cubic phase, giving a phase transformation temperature, $T_C$, of $\sim$ 512.5 K $\pm$ 12.5 K. This number is in fair agreement with experimental observations where $T_C$ has been reported at $\sim$ 440 K \cite{Zhang2018} or $\sim$ 480 K \cite{Maus_2023}; the difference likely being related to intrinsic defect concentration. The overestimation of $T_C$ is partly related to the overestimation of the $c/a$ of the r$^2$SCAN DFT functional used to train our MLFF (see Table S1). Another potential source of discrepancy is that we simulate pristine Na$_3$SbS$_4$, while the experimental samples contain intrinsic defects, in particular Na-vacancies.

Doping Na$_3$SbS$_4$ with W has been reported to drastically increase the Na diffusivity, as well as to decrease T$_C$ \cite{Maus_2023,Hayashi_2019,Fuchs_2019}. We start by investigating the reduction in T$_C$.  Fig.\ \ref{fig:phase_trans} (a) shows the averaged lattice parameters as a function of temperature of Na$_{3-x}$W$_x$Sb$_{1-x}$S$_4$ for W-dopant concentrations from 0 to 10\%. Note that we introduce one charge-compensating Na-vacancy for each W-dopant in the simulation cell. We observe a clear reduction of both $c/a$ in the tetragonal phase, and of $T_C$, with increasing W-content. We explicitly show $T_C$ as a function of dopant concentration in Fig.\ \ref{fig:phase_trans}, where we compare to experimental transition temperatures from XRD measurements by Maus \textit{et al.} \cite{Maus_2023}. We see that our results reproduce the experimental trend nicely, again with a slight overestimation. 
Another observation that can be made from Fig. \ref{fig:phase_trans} is that while W-doping has a large effect on the $c/a$ in the tetragonal phase ratio, and as a result T$_C$, its effect on the volume, once the systems have transformed into the cubic phase, is small.

\subsection{\label{sec:W_vs_Navac} Disentangling the effect of W-substitution and Na-vacancies}
\begin{figure}
\centering
\includegraphics[width=0.5\linewidth, keepaspectratio=true]{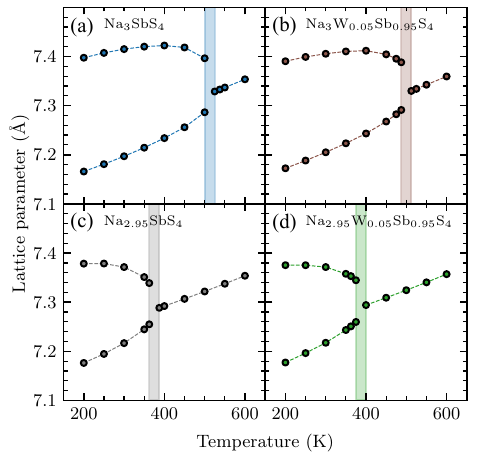}
\caption{Lattice parameters vs temperature of (a) pristine Na$_{3}$Sb$_{1}$S$_4$, (b) W substitution with no Na vacancies (Na$_{3}$W$_{0.05}$Sb$_{0.95}$S$_4$) (c) Na vacancies but no  W substitution (Na$_{2.95}$Sb$_{1}$S$_4$) and (d) W substitution with compensating Na vacancies (Na$_{3}$W$_{0.05}$Sb$_{0.95}$S$_4$). Shaded areas represent the phase transition region.}
\label{fig:phase_trans_W_vac}
\end{figure}

While experimental data has shown a trend of decrease in $T_C$ with increasing W-content \cite{Maus_2023}, which our MLMD simulations accurately reproduce (Fig.\ \ref{fig:phase_trans}), it is unknown whether this reduction is due to the W substitutions, the accompanying Na$^{+}$-vacancies, or a combined effect. To separate these influences, we perform NpT MD simulations, with the same setup as in Fig.\ \ref{fig:phase_trans}, for two additional systems: one with 5 \% W-dopants but no Na vacancies (Na$_{3}$W$_{0.05}$Sb$_{0.95}$S$_4$) and one with no W-dopants but an amount of Na-vacancies that corresponds to the charge-compensating amount for 5 \% W (Na$_{2.95}$SbS$_4$). 

In Fig.\ \ref{fig:phase_trans_W_vac} we compare these two systems to the 0 (Na$_{3}$Sb$_{1}$S$_4$), and 5 \%  (Na$_{2.95}$W$_{0.05}$Sb$_{0.95}$S$_4$) W-substituted ones from Fig. \ref{fig:phase_trans}. We can see that introducing Na vacancies has a large effect on the transition temperature (comparing Fig. \ref{fig:phase_trans} (a) and (c)), while introducing W-dopants without Na vacancies (comparing Fig. \ref{fig:phase_trans} (a) and (b)), has a negligible effect. Furthermore, the combined effect of W dopants and Na vacancies is small, as can be seen by comparing Fig.\ \ref{fig:phase_trans} (c) and (d). 

We may thus conclude that the reduction of $T_C$ in these systems by substitutional W-doping is an effect of the accompanying Na-vacancies, rather than the W-dopants themselves. This can be understood as follows: To retain the lower symmetry tetragonal phase, a modulation with a long correlation length must be maintained in the structure. Since the W-S bond length is only slightly shorter than the Sb-S one, W-doping (at these low concentrations) has a small effect on this modulation. On the other hand, introducing Na-vacancies, which are highly mobile, has a large destabilizing effect on this long-range modulation, and thus favours the cubic phase. 

These results may partly explain the observation that different sample preparation procedures have been observed to yield different phases (cubic or tetragonal) in Na$_3$PS$_4$ \cite{Krauskopf_2018}. Indeed, synthesis routes where relatively high concentrations of defects may be expected, tend to produce samples which are, on average, cubic \cite{Krauskopf_2018}.

\subsection{\label{sec:local_structure} Local vs Global Structure}
\begin{figure*} 
\centering
\includegraphics[width=\linewidth, keepaspectratio=true]{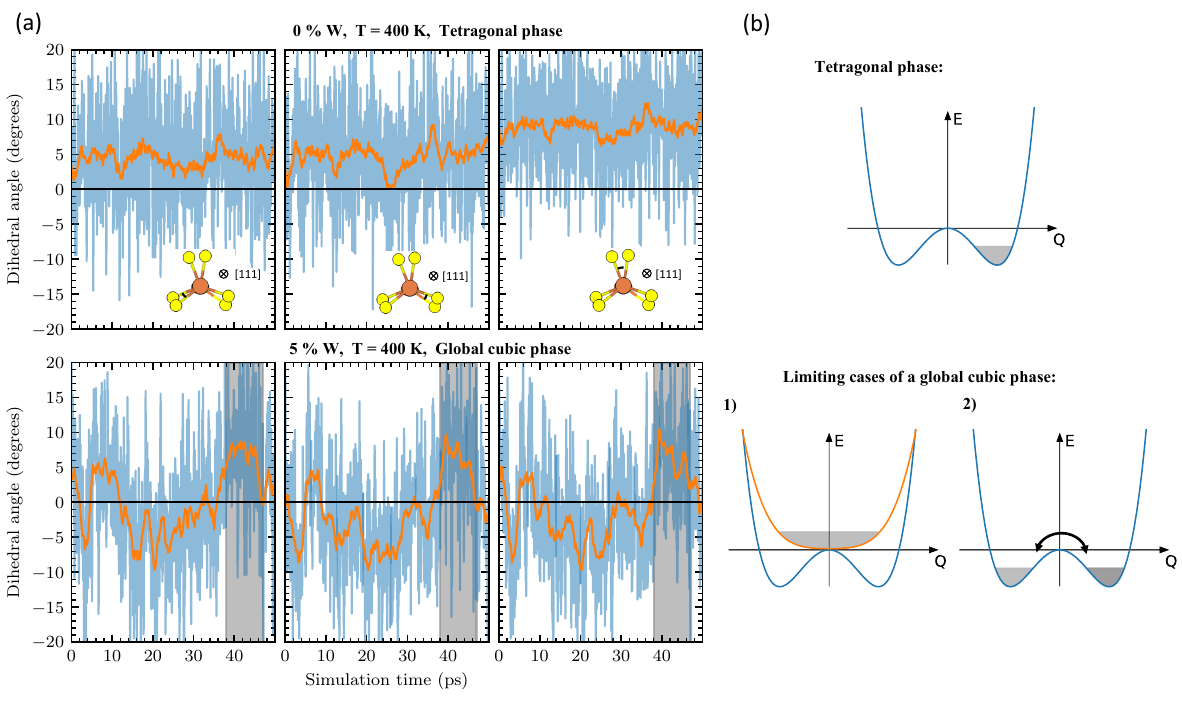}
\caption{(a) Dihedral angles associated with a selected sequential pair of SbS$_4$ tetrahedra along the [111] direction (illustration in the inset) as a function of MD simulation time. The top row shows the undoped system at 400 K (tetragonal phase), while the bottom shows the 5 \% W doped system at 400 K (cubic phase). The orange line corresponds to a square window average of 2 ps. The shaded region highlights one particular large and long lived deviation from the average cubic phase. (b) Cartoon schematic of the potential energy surfaces along a normal mode coordinate, $Q$, representative of the dihedral angle. $Q$ = 0 corresponds to high symmetry phase. Gray shaded area represents  the thermal population.}
\label{fig:diheds}
\end{figure*}

There have been several indications in the literature that this class of materials show deviations of their apparent average symmetry, as probed by eg. X-ray diffraction, and their local symmetry, as revealed by a local probe eg. PDF measurements \cite{Krauskopf_2018,Brenner2022,Maus_2023}. Indeed, Na$_{3-x}$W$_x$Sb$_{1-x}$S$_4$ shows an average cubic symmetry at high T, while being better described by the symmetry of the low T tetragonal phase on a local scale \cite{Maus_2023}. To probe this behaviour in our MD simulations we track the dihedral angles between successive SbS$_4$ tetrahedra along the [111] directions, illustrated in the inset of Fig. \ref{fig:diheds}. In the ideal cubic phase, the tetrahedra are perfectly aligned, resulting in dihedral angles equal to zero, while in the ideal tetragonal phase there is a relative tilt of the tetrahedra. 

The top row of Fig.\ \ref{fig:diheds} shows the 3 dihedral angles formed between a selected pair of consecutive SbS$_4$ tetrahedra along the [111] direction of pristine Na$_3$SbS$_4$ in the tetragonal phase at 400 K. We observe the expected behaviour where the dihedral angles oscillate around non zero values. Note that the tetragonal distortion results in one larger (right panel in the top row) and two smaller dihedral angles. 

The bottom row shows the 5 \% W  substituted system at 400K, which is above $T_C$, c.f. Fig.\ \ref{fig:phase_trans}. The expected behaviour in the cubic phase, where the ideal positions correspond to perfectly aligned tetrahedra, would be oscillatory motion of the dihedral angles around zero degrees, cf. Fig.\ \ref{fig:diheds} (b). We see that the long-time average value of the dihedral angles indeed is zero, consistent with the average cubic phase. It is also apparent, however, that there are large deviations from the cubic phase. Indeed, as exemplified by the shaded gray region, these deviations can exists over many ps. 

We note that for W substituted systems we have sequential pairs of both WS$_4$-SbS$_4$, SbS$_4$-SbS$_4$ and WS$_4$-WS$_4$ tetrahedra. We have found no qualitative difference between these. A set of random dihedral angles from several different sequential pair of tetrahedra of all types is shown in SI Fig 6. 

These results are in line with recent experimental observations of "local tetragonality" \cite{Krauskopf_2018,Brenner2022,Maus_2023} mentioned above. Indeed, when such large deviations from the global average symmetry occur at the local scale, it is expected that eg. a measured PDF fits better to the lower tetragonal symmetry over short distances.

The situation is schematically illustrated in Fig.\ \ref{fig:diheds} (b), where two limiting cases of a globally cubic phase is shown. The coordinate $Q$ is representative of the dihedral angle. In the first case, which corresponds to the "expected" behaviour mentioned above, there is an effective potential above $T_C$, and the system simply performs low frequency oscillations around $Q$=0. The other, limit is the case where the systems attains a cubic phase, i.e. $Q$=0, on average, through oscillations within an energy well overlaid with infrequenct stochastic hops between the wells. From Fig.\ \ref{fig:diheds} (a) we can recognize such hops for the 5 \% W substituted system in the cubic phase. 
 
\subsection{\label{sec:Diffusion}Diffusion}
\begin{figure}
\centering
\includegraphics[width=0.5\linewidth, keepaspectratio=true]{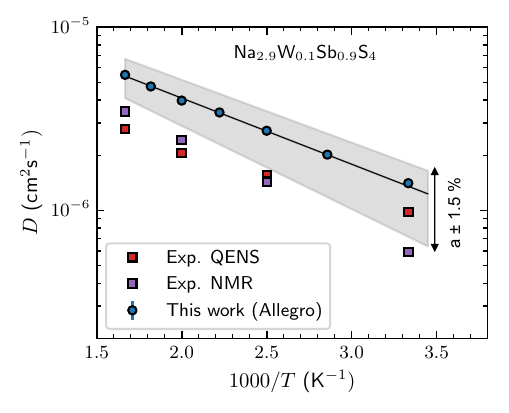}
\caption{Calculated self diffusion coefficient, $D$, of Na$_{2.9}$W$_{0.1}$Sb$_{0.9}$S$_4$ as a function of inverse temperature (blue circles), compared to experimental values extracted from QENS and NMR \cite{Maus_2023}. The black line represents an Arrhenius fit and the gray shaded area shows the sensitivity of the $D$ to a change in the lattice parameter, $a$, of $\pm$ 1.5 \%.}
\label{fig:diff_1}
\end{figure}

\begin{figure}
\centering
\includegraphics[width=0.5\linewidth, keepaspectratio=true]{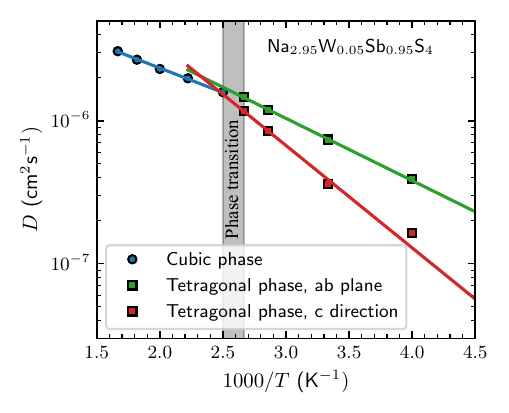}
\caption{Calculated self diffusion coefficient, $D$, of Na$_{2.95}$W$_{0.05}$Sb$_{0.95}$S$_4$ as a function of inverse temperature (blue circles), above and below the phase transition temperature, $T_C$ (shaded area). Below $T_C$, in the tetragonal phase, $D$ is calculated separately in the $c$ direction and in the $ab$-plane.  }
\label{fig:diff_2}
\end{figure}

Next we investigate how well our MLFF reproduces the diffusion behaviour. Fig.\ \ref{fig:diff_1} shows the calculated Arrhenius plot of the self diffusion coefficient, $D$, for Na$_{2.9}$W$_{0.1}$Sb$_{0.9}$S$_4$, compared to results extracted from quasi-elastic neutron scattering (QENS) and nuclear magnetic resonance (NMR) from Ref.\ \cite{Maus_2023}. We note a good agreement compared to experimental values, barring a slight overestimation. In particular, our calculated activation energy, corresponding to the slope of the Arrhenius fit, of $\sim$ 0.07 eV is in close agreement with the QENS and NMR values of $\sim$ 0.05 and 0.09 eV, respectively. The overestimation of $D$ is partly related to the overestimation by r$^2$SCAN of the volume. To highlight the sensitivity of the calculated diffusion coefficients to the volume, we show results using lattice constants changed $\pm$ 1.5 \%, highlighted in gray in Fig.\ \ref{fig:diff_1}. 

At this point, we can ask the question to what extent the W dopants influence the self diffusion, beyond inducing Na-vacancies. By comparing diffusion coefficients between calculating with and without W dopants, but with the same number of Na-vacancies, we find that the W dopants have a minor detrimental effect on the predicted diffusivities (see SI Fig.\ 7). We may thus conclude that W dopants have a weak explicit effect on both the structural and phase behaviour, and on the ionic diffusion, and their utility in these systems is limited to a means of inducing Na-vacancies.

Next we probe the effect the phase transformation on the diffusion. Fig. \ref{fig:diff_2}  shows the diffusion coefficient of Na$_{2.95}$W$_{0.05}$Sb$_{0.95}$S$_4$ in the cubic and tetragonal phases. In the tetragonal phase ,we separately calculate the diffusion coefficient along the $c$ direction and in the $ab$ plane and find a significant anisotropy. Indeed, the activation energy for diffusion in the $c$ direction is $\sim$ 0.14 eV, significantly larger than $\sim$ 0.09 eV, in the $ab$ plane.

The general connection between vibrations and diffusion in potential solid-state electrolyte materials has been studied intensively in recent years \cite{Bachman2015,Krauskopf2017}. For the present class of materials in particular, Gupta \textit{et al.} \cite{Gupta2021} proposed a connection between the soft phonon mode (see Fig.\ \ref{fig:fig1}) and the ionic diffusion. Indeed, it is easy to imagine that a low energy phonon, indicative of a shallow potential energy surface, with eigendisplacements aligning closely with the Na migration path would be beneficial for diffusion. Furthermore, since the soft phonon mode (see Fig.\ \ref{fig:fig1}) contains concerted motion of whole chains of Na ion along a [100] direction, one would expect diffusion events triggered by this phonon mode to involve several Na ions. From animations of our MD trajectories (see Suppl. video), it can indeed be appreciated that the diffusion process does not consist solely of isolated Na hops, but that there is typically concerted motion of several Na ions along the diffusion pathways. To quantify this behaviour we calculate the Haven ratio, $H_R$, which is a measure of the degree of correlation in the diffusion process of a material (see methods). Loosely speaking, $H_R <$ 1 indicates that distinct Na ions tend to preferentially move in the same direction, as would be expected in our case. For Na$_{2.9}$W$_{0.1}$Sb$_{0.9}$S$_4$ we find values $\sim$ 0.35 $\pm$ 0.1 in the whole temperature range of the cubic phase (see SI Fig.\ 8). We note that converging $H_R$ requires significantly more statistics than the self diffusion coefficient, and we don't have enough data resolve any potential temperature dependence. Nevertheless, these values of $H_R$ indicate significant concerted diffusion and are in the range of values found in superionic conductors, eg. Li$_3$N \cite{Messer1981,Krenzer2023} and Bi$_2$O$_3$ \cite{Mohn2020}.

\section{\label{sec:conclusion} Conclusions} 
We have constructed a machine learning force field capable of describing the phase transformations and diffusion over a range of substitutional W doping in Na$_{3}$Sb$_{1}$S$_4$, a promising Na-ion electrolyte material. We reproduce the experimentally known trend of decreasing tetragonal-to-cubic phase transition temperature on increasing W-content and reveal that this reduction is an effect of the charge-compensating Na-vacancies, rather than the W-dopant themselves. We further show that the cubic phase displays large local deviations from the average symmetry, in line with recent experimental suggestions. Our MLFF further reproduces, barring a slight overestimation, the temperature dependence of the experimental self diffusion coefficient, and again reveals that the explicit effect of W substitution is small, and that the diffusion behaviour is governed by the Na-vacancies. Our work shows that carefully constructed force fields, using modern architectures, can describe the highly complex interplay between substitutional doping, structural phase transformations and diffusion in promising Na-ion solid state electrolytes.

\begin{acknowledgement}

J. K. acknowledges support from the Swedish Research Council (VR) program 2021-00486. Computations were enabled by resources provided by the National Academic Infrastructure for Supercomputing in Sweden (NAISS) at NSC and PDC partially funded by the Swedish Research Council through grant agreement no. 2022-06725. The training of the machine-learned force fields were enabled by the Berzelius resource provided by the Knut and Alice Wallenberg Foundation at the National Supercomputer Centre. We also acknowledge the National Academic Infrastructure for Supercomputing in Sweden (NAISS) partially funded by the Swedish Research Council through grant agreement no. 2022-06725 for awarding this project access to the LUMI supercomputer, owned by the EuroHPC Joint Undertaking, hosted by CSC (Finland) and the LUMI consortium. Via our membership of the UK’s HEC Materials Chemistry Consortium, which is funded by EPSRC (EP/X035859/1), this work used the ARCHER2 UK National Supercomputing Service (http://www.archer2.ac.uk). A. W. acknowledges support from EPSRC project EP/X037754/1.

\end{acknowledgement}

\begin{suppinfo}

The Supporting information contains additional validation of the MLFF, and additional figures relating to the phase transformation and  the diffusion process.
and a video illustrating the diffusion process for 5 \% W-doped \ce{Na3SbS4} at 450 K. This material is available free of charge via the Internet at
http://pubs.acs.org.

%This will usually read something like: ``Experimental procedures and
%characterization data for all new compounds. The class will
%automatically add a sentence pointing to the information on-line:

\end{suppinfo}

%\bibliography{main}
\providecommand{\latin}[1]{#1}
\makeatletter
\providecommand{\doi}
  {\begingroup\let\do\@makeother\dospecials
  \catcode`\{=1 \catcode`\}=2 \doi@aux}
\providecommand{\doi@aux}[1]{\endgroup\texttt{#1}}
\makeatother
\providecommand*\mcitethebibliography{\thebibliography}
\csname @ifundefined\endcsname{endmcitethebibliography}
  {\let\endmcitethebibliography\endthebibliography}{}

\end{document}